\magnification=1200

\font\eightrm=cmr8

\input psfig.sty
\centerline{\bf FERMION GENERATIONS FROM THE HIGGS SECTOR}

\vskip 1.0cm
\centerline{Vladimir Visnjic}
\centerline{\it Department of Physics, Temple University,
Philadelphia, PA 19122}
\centerline{and}
\centerline{\it Institut za Fiziku, Beograd, Yugoslavia}
\centerline{E-mail: visnjic@astro.temple.edu}

\vskip 1.0cm

\centerline{\bf Abstract}

\vskip 4mm

{\eightrm
The generation structure in the quark and lepton spectrum is
explained as originating from the excitation spectrum $S_n$ 
of SU(2)$_W$ doublet scalar fields, whose ground state $S_1$ 
is the Standard Model Higgs field. There is only one basic 
family of SU(2)$_W$ doublet left-handed fermions,
$\nu_L,e_L,u_L,d_L$, whose bound states with  
$S_n$ manifest themselves as the generations of left-handed 
quarks and leptons. Likewise, there is only one basic family 
of the right-handed
fermions, $\nu_R,e_R,u_R,d_R$, which combine with the 
gauge invariant scalar fields $G_n$ to produce the right-handed 
quarks and leptons of the second and higher generations. There
are only four Yukawa coupling constants, $G_\nu,G_e,G_u$, and
$G_d$ and all quark and lepton masses are proportional to them.
Suppression of flavor changing neutral currents (GIM mechanism) 
is automatic. $\nu_\mu$ and $\nu_\tau$ are expected to be massive.}

\vskip 5mm

I present a theory of the origin of fermion generations
in which there is only one fundamental quark/lepton family, while 
the second and higher ones are a consequence of an excitation 
spectrum in the
scalar sector. The basic family of chiral fermions consists of
$$\ell_L=\pmatrix{\nu_L\cr e_L},\,
q_L=\pmatrix{u_L\cr d_L},\nu_R,e_R,u_R,d_R$$
with the usual
${\rm SU(2)}_W\otimes {\rm U(1)}_Y$ transformation
properties. In addition to the fermions, there is a composite bosonic 
field $S$ whose lowest energy states are the SU(2)$_W$ doublet scalars
$$ S_1=\pmatrix{S_1^+\cr S_1^0},\,S_2=\pmatrix{S_2^+\cr S_2^0},\,
S_3=\pmatrix{S_3^+\cr S_3^0}\dots$$
We shall assume that there are at least three scalar states  
below the lowest $J=1$ state. The ground state $S_1$ is the familiar 
Higgs field which develops a vacuum expectation value. $S_2,S_3,$ etc. 
are its radial excitations 
labeled by the ``generation number.'' Orthogonality of the states
$S_1,S_2,S_3\dots$ implies that for $n\geq 2$, $S_n$ has 
zero vacuum expectation value and zero Yukawa couplings.
It also requires that the effective quartic couplings of these states
respect the generation number, at least at energies much lower
than their mass scale. Under this assumption the
effective potential can only be a function of
$S^\dagger _iS_i$ and $(S^\dagger _iS_j)(S^\dagger _jS_i)$
and the low energy effective theory possesses a global SU(2)
symmetry, the isospin.

Note that this scheme has nothing to do with Technicolor which is a
QCD-like theory invented to solve the fine-tuning problem and 
does not address the fermion generation problem. The $S$-field
envisaged here explains the existence of fermion generations as its
radial excitation states. For this purpose it is not necessary
to assume QCD-like structure for the $S$-field -- or even 
that it is based on a new interaction at all.

\vskip 3mm

{\bf Gauge invariant scalars}

\vskip 3mm

The gauge invariant scalar fields are obtained in the $S_i^\dagger S_j$
(neutral) and $\bar S_i^\dagger S_j$ (charged) channels, where
$\bar S=i\tau_2S^*$. The fields
involving the ground state $S_1$ are of particular importance,
since $S_1$ is the only Higgs field in the theory (i.e. the field
which has negative mass squared in the Lagrangian) and plays a
special role
in producing the $W$'s, the $Z$, and the left-handed fermions of
the first generation. The only gauge invariant scalar involving
only $S_1$ is $h^0=\sqrt{2}({S_1^0-<S_1^0>)}$, the
physical Higgs field. For every $n\geq 2$ we have a positive 
scalar $G^+_n=\bar S_1^\dagger S_n$ and a neutral scalar 
$G^0_n=S_1^\dagger S_n$, both of which are 
labeled by $n$ and thus come in generations. As we
shall see, the $G_n$ fields give their generation number to the
right-handed fermions and thus are responsible for the generation
structure in the right-handed sector.

\vskip 3mm

{\bf The quarks and leptons}

\vskip 3mm

The left-handed fermions of the $n$-th generation,
$\nu_{nL},e_{nL},u_{nL},d_{nL},$ are composed of the fundamental
left-handed 
fermions $\ell_L$ and $q_L$ and the $n$-th generation $S$ field, $S_n$:
$$\eqalign {\nu_{n L}&=\bar S_n^\dagger\ell_L\cr
                       e_{nL}&=S_n^\dagger\ell_L\cr
                       u_{nL}&= \bar S_n^\dagger q_L\cr
                       d_{nL}&= S_n^\dagger q_L.\cr}\eqno(1)$$
The right-handed fermions of the $n$-th generation are composed
of the fundamental right-handed fermions $\nu_R,e_R,u_R,d_R,$ and the 
$G_n$ fields from which they get their generation label:
$$\eqalign {\nu_{n R}&=G_n^+e_R+G_n^0\nu_R\cr
                       e_{nR}&=G_n^{0*}e_R+G_n^-\nu_R\cr
                       u_{nR}&=G_n^+d_R+G_n^0u_R\cr
                       d_{nR}&=G_n^{0*}d_R+G_n^-u_R.\cr}\eqno(2)$$
The left- and the right-handed fermions of the first generation meet at
the usual Yukawa vertices, Fig. 1(a), while those of the second and higher
generations meet at the vertices shown in Fig. 1(b). The four Yukawa 
coupling constants, $G_\nu,G_e,G_u$, and $G_d$ are the only chiral
symmetry 
breaking parameters in the theory and thus all fermion masses are 
proportional to them. In the limit in which these coupling constants 
go to zero all fermions are massless, irrespective of how
massive their scalar constituents $S_i$ may be. Note, however, that the
quark (lepton) masses of the second and higher generations get
contributions from {\it both} $G_u$ and $G_d$ ($G_\nu$ and $G_e$),
$$\eqalign{m_c&=A m_u+B m_d\cr
m_s&=C m_u+D m_d\cr}$$
and analogous in the lepton sector. In particular, this implies
that the muon and tau neutrinos are massive even if $G_\nu=0$,
due to the contributions from the electron to those two neutrinos.

\bigskip

\psfig{file=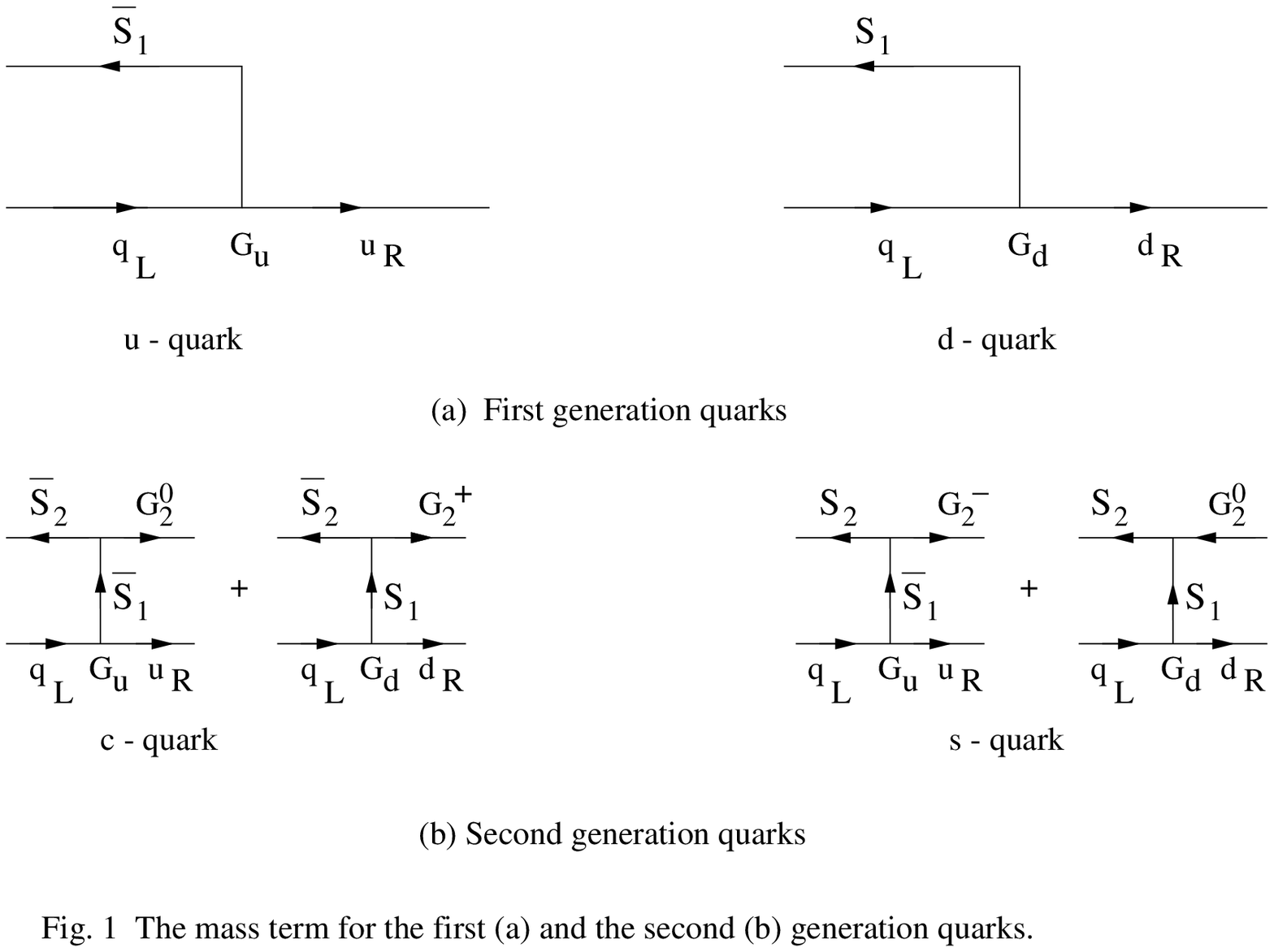,width=5.3in}

\vskip 1cm

{\bf The W and Z couplings and the mixing angles}

\vskip 3mm

In the Standard Model the $W$ and $Z$ bosons couple to 
$\bar S_1^\dagger (A_\mu-{i\over g}\partial_\mu ) S_1$ and
$S_1^\dagger (A_\mu-{i\over g}\partial_\mu ) S_1$ where $A_\mu$ 
denotes the SU(2)$_W$ gauge field. Here I postulate similar
couplings of the $W$ and $Z$ bosons to the excited states, i.e. to
$\bar S_i^\dagger (A_\mu-{i\over g}\partial_\mu )S_j$ and
$S_i^\dagger (A_\mu-{i\over g}\partial_\mu)S_j$. 
This makes it possible to couple the $W$'s and the $Z$ to the 
quarks and leptons. However, the quark and lepton mass eigenstates are
the bound states, Eq. (1). The $W$ coupling to the mass
eigenstates involves the mixing matrix elements $V_{ij}$, equal
to the overlap integral of the quark wave functions,
$$V_{ij}=<u_{iL}|d_{jL}>=<\bar S_i^\dagger q_L|S_j^\dagger q_L>.$$
As shown in Ref. 1, orthogonality 
and completeness of the quark wave functions ensure the unitarity of 
the mixing matrix $V$ and thus the absence of flavor changing neutral 
currents (GIM mechanism). Alternatively, we may directly 
show that the off-diagonal couplings of the $Z$ involving $S$ 
fields of different generations are zero because of orthogonality 
of the $S_n$ states. Higher
order corrections, of course, do not respect the orthogonality and
introduce flavor changing neutral currents.

\vskip 3mm

{\bf Conclusions}

\vskip 3mm

The proposed theory of quark and lepton generations predicts that,
except for the first one, each quark and lepton generation is 
accompanied by a pair of scalar fields, $G_n^+$ and $G_n^0$. Thus 
both the fermions and the scalars
come in generations which originate in the spectrum of the
$S$-field. Ultimately the model should predict the number 
of quark/lepton generations and their masses. Even in the absence
of these predictions, we may notice that the fermion generations
predicted here differ in two important aspects from higher
fermion generations which would obey the rules of the present Standard
Model. First, although they may be heavy, they do not involve
large Yukawa couplings and thus do not lead to new strong
interactions among quarks/leptons. Heavy fermions are obtained
as bound states and not as a result of large Yukawa couplings. Second, 
even if the number of quark generations turns out to be large,
the asymptotic freedom properties of QCD may not be affected, since 
there are only two basic quark flavors, $u$
and $d$, all other quark flavors being bound states of these two.

\vskip 1cm

\centerline{\bf References}

\medskip

\noindent
1. V. Visnjic, {\sl Phys. Rev.} {\bf D25} (1982) 248.

\vfill \eject

\end

According to the view presented in this work, the Standard Model Higgs
field and the existence of fermion generations are but two tips of
the same iceberg. Hidden beneath the surface lies a huge body of
new physics.

\end